# Peierls-type metal-insulator transition in carbon nanostructures

*Bing Zhang[a], Ting Zhang[a], Jie Pan[a], Tsz Pong Chow[a], Ammar M. Aboalsaud[b], Zhiping Lai[b,*], Ping Sheng[a,†]*

[a]*Department of Physics, Hong Kong University of Science and Technology, Clear Water Bay, Kowloon, Hong Kong, China*

[b]*Division of Physical Sciences and Engineering, King Abdullah University of Science and Technology (KAUST), Thuwal 23955-6900, Saudi Arabia*

**Abstract**

We report the observation of Peierls-type metal-insulator transition in carbon nanostructures formed by chemical vapor deposition inside the pore network of the ZSM-5 zeolite. The Raman spectrum of this nanocarbon@ZSM-5 indicates a clear signature of the radial breathing mode (RBM) for (3,0) carbon nanotubes that can constitute the carbon network segments. Electrical transport measurements on multiple few-micron-sized nanocarbon@ZSM-5 crystals showed metallic temperature of resistance dependence down to 30 K, at which point the resistance exhibited a sharp upturn that is accompanied by the opening of a quasigap at the Fermi level as indicated by the differential resistance measurements. Further Hall measurements have yielded both the sign of the charge carrier and its density. The latter demonstrated excellent consistency with the quasigap data. We employed first-principles calculations to verify that there can indeed be softening of the phonon modes in the (3,0) carbon nanotubes.

---

[*] Corresponding author. E-mail: zhiping.lai@kaust.edu.sa (Zhiping Lai)

[†] Corresponding author. E-mail: sheng@ust.hk (Ping Sheng)

## 1. Introduction

Zeolites are microporous crystalline aluminosilicates that have uniform porous structures. The pore diameters are in the range between 0.3 nm and 1.2 nm. The robust porous skeletal structure of zeolites has been widely used as catalysts, adsorbents, porous supports, and hosts for making nanomaterials [1, 2]. In this work we use the calcined ZSM-5, a type of 10 member-ring zeolite with framework code MFI [3-5], as a template for forming carbon nanostrucutres in its pore network by using the chemical vapor deposition (CVD) method [6]. The skeletal structure of ZSM-5 is shown below in Fig. 1(a); it has straight channels along the b-axis and sinusoidal channels along the a-axis. Both channels have an inner diameter of ~5 Angstroms and interconnected with each other with a segment distance ~ 10 angstroms. Each ZSM-5 crystal is about 2 to 3 microns across, as shown in Fig. 1(b). In contrast to previous works in which 4 Angstrom carbon nanotubes [7-9] were formed inside the linear channels of AFI zeolite [6, 10, 11], here the nanocarbon@ZSM-5 has a three dimensional (3D) structure. Characterization by Raman spectroscopy indicates a clear radial breathing modes (RBM) peak at 805 $cm^{-1}$ that agrees extremely well with that for the (3,0) carbon nanotube with a diameter of 3 Angstroms. Four-probe electrical resistance measurement on a single zeolite crystal showed a metallic temperature dependence down to 30 K, with a sharp reversal below that temperature. Differential resistance measurements below 30 K showed the development of a quasigap at the Fermi level, which is a signature of the Peierls-type metal-insulator (M-I) transition [12-14]. This type of M-I transition differs from the more common types of M-I transitions in carbon materials such as those induced by molecular doping in graphene [15], by redox doping in single-wall carbon nanotubes [16], or as that in iodinated amorphous carbon films [17]. In addition, charge carriers localization can also lead to M-I transition, e.g., in highly disordered carbon fibers [18, 19]. To our knowledge, this is the first experimental observation of Peierls-type transition in carbon nanomaterials. To confirm our interpretation, we have carried out Hall measurements on a single ZSM-5 crystal and obtained the variation of charge carrier density as a function of temperature below 30

K. The Fermi level carrier density showed a sharp decrease below 30 K, in excellent agreement with the differential resistance quasigap data.

Theoretical research on Peierls transition was widely studied in one-dimensional carbon materials, such as on single-wall carbon nanotubes CNT(5,5), CNT(3,3), CNT(5,0) [20-24], or in linear carbon chains [25]. In support of our experimental results, we have carried out first-principles calculations on (3,0) nanotubes and showed the existence of phonon soft modes as a result of electron-phonon coupling.

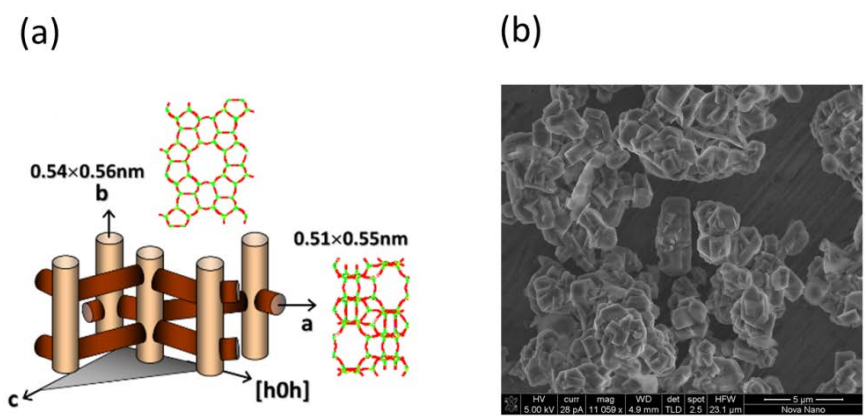

Figure 1. (a) A cartoon picture of ZSM-5's skeletal pore structure. (b) A SEM image of the ZSM-5's crystals. Each crystal is about 2 to 3 microns across.

2. Experimental

The nanocarbon@ZSM-5 was grown by using the CVD method. The calcined ZSM-5 zeolite crystals were put in a quartz tube. Five atmosphere pressure of methane ($CH_4$) was heated to 800℃ for 10 hours to grow the carbon nanostructures in the pores of ZSM-5. Subsequent to cooling down, nanocarbon@ZSM-5 was characterized by both Raman spectra and thermal gravimetric analysis (TGA). Results are detailed in the following section.

To measure the transport properties of nanocarbon@ZSM-5, we first deposit a thin layer of photoresist (950 PMMA 9 A) on a quartz substrate. Nanocarbon@ZSM-5 crystals were dispersed on the PMMA and heated on a hotplate at 180℃ for 90 seconds. This process was helpful to fix the dispersed crystals on the quartz substrate for the subsequent processing. A layer of adhesive Ti with a thickness of 5 nm was coated on the crystals by sputtering, followed by a layer of Au with a thickness of 60 nm. Focused ion beam (FIB) was used to select one crystal, on which the Ti/Au film was etched into a designed square geometric pattern as shown schematically in Fig. 2(a) with the four numbered electrical leads. A scanning electron microscope (SEM) image of the actual device is shown in Fig. 2(b). Due to the small size of the crystals, the traditional six-lead Hall bar geometry was not possible. To measure the longitudinal resistance, current was passed between leads 1 and 2 as shown in Fig. 2(a), and voltage was measured across leads 3 and 4. For the Hall measurements, current was passed between leads 1 and 3 under an applied magnetic field, and voltage was measured across leads 2 and 4.

Measurements of the fabricated device were carried out by using the Physical Property Measurement System (PPMS). A Keithley 6221 was used as the current source and a SR850 lock-in was used as voltmeter to measure the resistance and differential resistance of nanocarbon@ZSM-5.

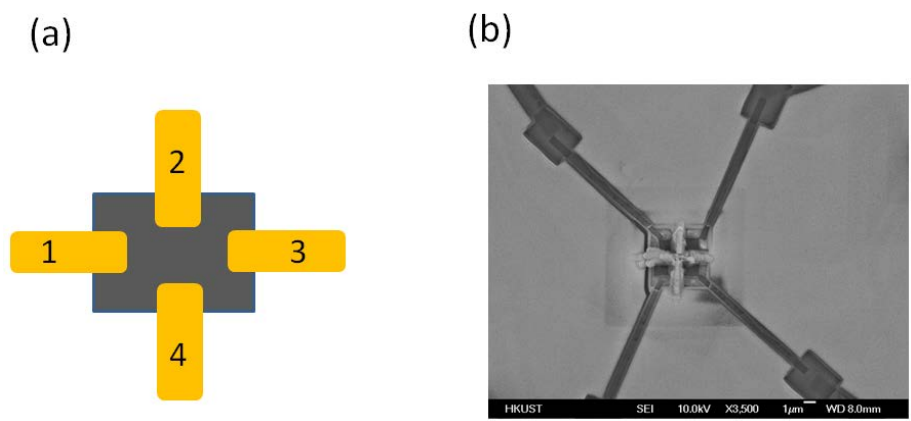

**Figure 2. Device for transport measurement. (a) Schematic illustration of the four-terminal geometry. (b) A SEM image of the four-terminal device for electrical measurement. The size of the nanocarbon@ZSM-5 sample is about 4 microns.**

3. **Results and discussion**

Measured Raman spectrum is shown in Fig. 3(a). The very high peak at 1604 cm$^{-1}$ is the G band which arises from the vibrations of C-C sp2 bonds. The peak at 1388 cm$^{-1}$ is the D band which arises from the defects in the structure of nanocarbon@ZSM-5. A lower D band peak usually indicates a lower concentration of defects in the sample. We have tried to adjust the fabrication conditions to minimize the defects. The peak at 805 cm$^{-1}$ is the radial breathing modes (RBM) of carbon nanotubes. It is well known that the RBM originates from the coherent vibration of the carbon atoms along the radial direction, which is unique to carbon nanotubes. Normally $\omega_{RBM} = \frac{A}{d} + B$, where $A$=228, $B$=16, with $d$ the nanotube diameter in units of nm. Typical RBM range is 100–350 cm$^{-1}$. From the measured Raman breathing mode data we deduce $d$=0.29 nm, indicating the nanocarbon@ZSM-5 structure comprises a network of (3,0) carbon nanotubes. From the ab initio calculations as that described in the Supplemental Materials, the RBM frequency of the (3,0) carbon nanotube is around 817 cm$^{-1}$, very close to the experimentally observed value of 805 cm$^{-1}$. The RBM of the (2,1) carbon nanotube, 0.236 nm in diameter, is noted to be on the order of 10% higher in frequency. Hence the identification of (3,0) carbon nanotube from its RBM is rather unique. The peak at 1192 cm$^{-1}$ can be due to the combination of RBM and D band modes, usually denoted the intermediate frequency modes (IFM) [26, 27] that are composed of both first and second-order modes. For comparison, the Raman spectrum for the empty ZSM-5 template is also shown in Fig. 3(a) as the blue curve. It is seen that there are no visible peaks in the relevant frequency range. This comparison indicates that the Raman signal indicated by the black curve in Fig. 3(a) is from the nano-carbon structure.

Results of the TGA measurements on crystals of nanocarbon@ZSM-5 are shown in Fig. 3(b). The sample was placed in the TGA equipment Q5000 and heated in air from room temperature to 800℃ with a heating rate of 2℃/min. The sample weight was monitored by a microbalance. By burning off the carbon inside the ZSM-5 crystals, we obtained the carbon's weight content from the difference in weight before and during the heating process. As shown in Fig. 2(b) the carbon content of nanocarbon@ZSM-5 is 14.3wt%. By assuming the carbon structure to be (3,0) carbon nanotubes in each segment of the structure, this TGA result translates into a pore occupation ratio of 35%. Detailed calculation that leads to this number is given in the Supplemental Materials, Section A. The derivative of the weight curve (blue) shows a sharp peak at around 600 ℃, which indicates the decomposition temperature of nanocarbon structure inside the pores of ZSM-5.

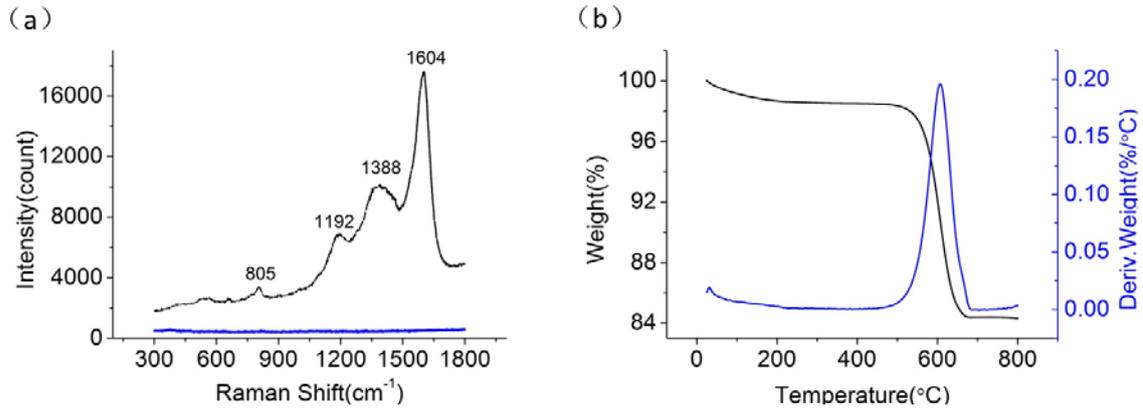

**Figure 3. (a) The Raman spectra of nano-carbon@ZSM-5 (black curve) and empty ZSM-5 template (blue curve), the latter showing no signal, and (b) the thermogravimetric analysis (TGA) of the nano-carbon@ZSM-5 formed by the CVD process. The TGA result shows that the carbon content from the difference in weight before and after the heating process is 14.3wt%. The blue curve is the differential of the weight loss curve, showing that 600 ºC is the decomposition temperature of nanocarbons inside the zeolite pores.**

Figure 4(a) shows the temperature variation of the measured longitudinal resistance in nanocarbon@ZSM-5. Very good linear variation of the resistance, characteristic of metal, is seen from room temperature down to 30 K. However, when the temperature was lowered below 30 K, the resistance is seen to increase quickly. Differential resistance measurements at four temperatures, plotted as a function of the driving current, is shown in Fig. 4(b). It shows that the mechanism of the resistance upturn below 30 K is associated with the development of a quasigap centered at the Fermi level. It is seen that at 30 K, the flat differential resistance indicates a linear, Ohmic I-V behavior. However, as the temperature is lowered a clear nonlinear I-V behavior is seen that can be interpreted as the development of a quasigap with an increasing resistance at the Fermi level due to the depletion of charge carriers. The magnitude of the resistance upturn, however, is relatively small, which means that only part of the sample has undergone a Peierls transition. Therefore we should consider the measured resistance to comprise a background resistance that follows the downward linear trend as a function of temperature, shown by the red dashed line in Fig. 4(a), that is in series with the part that has undergone a Peierls transition.

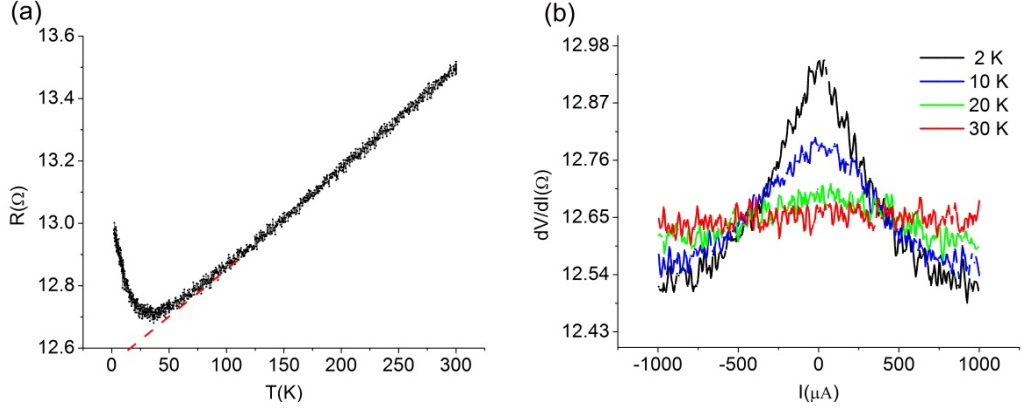

**Figure 4.** (a) Measured temperature dependence of resistance for nano-carbon@ZSM-5. A sharp resistance upturn is seen below 30 K. The red dashed line indicates a metallic resistance in series with the Peierls transition component. (b) Measured differential resistance plotted as a function of the driving current. The appearance of a sharp peak at the Fermi level is indicative of the development of a quasigap driven by the soft phonon mode, i.e., a Peierls transition. The sample was measured by the four-terminal configuration shown in Fig. 2(a).

In order to substantiate our interpretation that the upturn in resistance below 30 K is due to the depletion of charge carrier density at the Fermi level, we have carried out Hall measurements under an applied magnetic field. The knowledge of *both the longitudinal and transverse resistance would enable us to separately* determine the mobility and charge carrier density. The results are shown in Fig. 5(a). We know that the Hall resistivity $\rho_{xy} = \frac{B}{ne}$, where $B$ is the magnetic field, $n$ denotes the charge carrier density, and $e$ is the electronic charge. The slope of the Hall resistance is seen to increase with decreasing temperature, indicating a decrease in the charge carrier density at the Fermi level as the temperature is lowered. The data are plotted as black symbols in Fig. 5(b).

Since the longitudinal resistivity is given by $\rho_{xx} = \frac{1}{ne\mu}$, where $n$ denotes the charge carrier density and $\mu$ the mobility. If we consider the sample as a Peierls transition component in series with a metallic series resistance (red dashed line in Fig. 4(a)), then we have $\rho_{xx} = \frac{1}{ne\mu} + (aT + b)$, where $aT + b$ is the linear metallic resistance in series as shown by the (extrapolated) red dashed line in Fig. 4(a), with $a = $ 3.23 × $10^{-3}$ Ω T$^{-1}$ and $b = 10.80$ Ω. By fitting the longitudinal resistance data at temperatures higher than 30 K, we obtain the mobility value to be $\mu = 13.06$ $cm^2$ $V^{-1}s^{-1}$, which is lower than that in the plane of graphite, $\mu = 9.0 \times 10^4$ $cm^2$ $V^{-1}s^{-1}$ [28]. By assuming the mobility value to be a constant over the relevant temperature range, the value of *n* can be directly obtained from the peak value of the differential resistance data in Fig. 4(b) (as it corresponds to the Fermi level). The charge carrier density obtained in this manner is plotted as red symbols in Fig. 5(b). We can see that the agreement is excellent, which not only confirms our assumption that the mobility is constant over the relevant temperature range, but also leads to the conclusion that a Peierls-type M-I transition has indeed occurred through the opening of a quasigap in the Fermi level charge carrier density.

It should be noted that such resistance upturn has been observed on multiple samples of nanocarbon@ZSM-5. Therefore this phenomenon is robust.

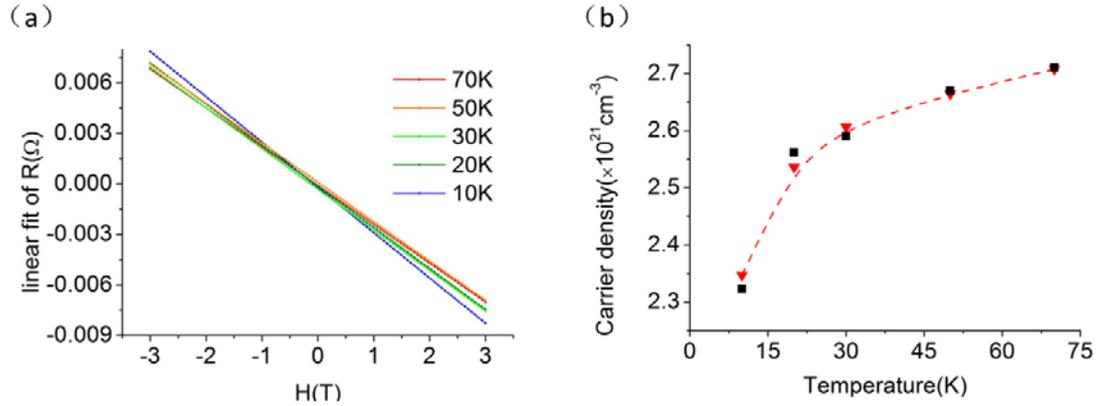

**Figure 5. (a) The temperature dependence of Hall resistance for nano-carbon@ZSM-5. (b) A comparison between the electronic charge carrier density obtained from by the Hall measurements (black symbols) and that obtained from fitting of the longitudinal differential resistance peak values vs. temperature in nano-carbon@ZSM-5 (red symbols). The red dashed curve is to guide the eye.**

## 4. Theoretical considerations

Peierls-type M-I transition owes its origin in electron-phonon coupling and the one-dimensional geometry. Since the ZSM-5 zeolite pores comprise intersecting one-dimensional channels with 5 Angstrom inner diameter, (3,0) CNT can be accommodated. Here we would like to consider the nanocarbon@ZSM-5 to consist of (3,0) CNT segments, which is supported by the Raman RBM signal at 805 cm$^{-1}$. This value is very close to the ab-initio estimation of the RBM for the (3,0) CNT at 817 cm$^{-1}$. Our theoretical consideration is to see if the (3,0) CNT can have phonon softening, which is a necessary condition (but may not be sufficient) for the nanocarbon@ZSM-5 to exhibit a Peierls-type M-I transition [12, 13].

In a Peierls process the lattice distorts (usually dimerizes) and opens a gap at the Fermi level, thereby makes the system transit from a metal to an insulator at zero temperature. In such a process the equilibrium positions of atoms change, causing an increase in the elastic energy. However, distortion of the atomic lattice would simultaneously open a gap in the electronic band structure at the Fermi level and lowers the electronic energy. As the latter more than compensates the increase in the elastic energy, hence this is an instability that can occur with especially high inevitability in 1D metals. However, if the atomic bonding is strong, the Peierls transition temperature can be low. We have carried out ab-inito calculations on the (3,0) CNTs. The calculational details are given in the Supplemental Materials, Section B. Our estimation shows that the (3,0) CNT can indeed have phonon softening that undergoes a Peierls transition under $T_\mathrm{p}$ ~273 K. The relevant soft phonon mode of (3,0) CNT is illustrated in Fig. 6. We have to recognize, though, that the present nanocarbon@ZSM-5 is a 3D network of 1D segments, hence the theory necessarily over-estimate the Peierls transition temperature.

The existence of the Peierls-type transition in nanocarbon@ZSM-5 implies a reasonably large electron-phonon coupling. Since superconductivity is another consequence of electron-phonon coupling, in our earlier theoretical works [29, 30] we have found that better screening of Coulomb interaction among the electrons, which is a negative for superconductivity, can greatly suppress the Peierls transition while simultaneously enhance the superconductivity transition. As a result, our next experimental effort is along the direction of providing better Coulomb screening, with the intention to see if a superconducting ground state can be accomplished in this system.

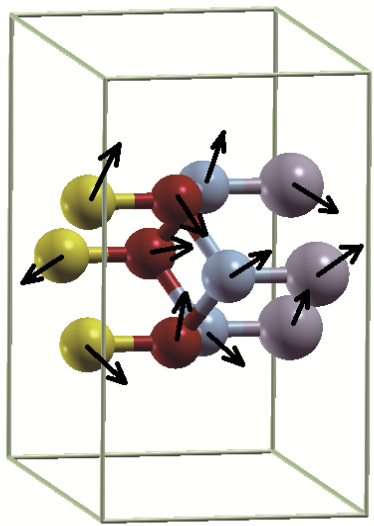

**Figure 6. The unit cell of (3,0) CNT consists of four triangular layers of carbon atoms, with two registered layers in the back rotated 60 degrees with respect to the two registered layers in the front. Here each layer is identified with a different color from the other layers, to facilitate easy visualization. The arrows shown in the figure indicate the magnitude and direction of the displacement for each atom in the soft phonon mode with q=0.45 / *a*, where *a* is the length of the unit cell.**

5. Conclusions

By making use of the ZSM-5 zeolite as the template we have successfully fabricated nano-carbon@ZSM-5 with a high pore filling factor (70%). The Raman spectrum of nanocarbon@ZSM-5 exhibits an RBM peaks around 805 cm$^{-1}$, indicating the nanocarbon@ZSM-5 to comprise segments of the smallest carbon nanotube (3,0). The resistance of the nanocarbon@ZSM-5 displays a metal to insulator transition at 30 K. This interesting phenomenon can be explained by the Peierls-type transition that opens a gap at the Fermi level. By carrying out resistance and Hall measurements we found consistency between the Hall measurement and the differential resistance measurement that clearly indicate the opening of a quasigap at the Fermi level. First-principles calculations on (3,0) CNT were carried out, with the results showing a Peierls transition can indeed occur.

**Declaration of competing interest**

The authors declare that they have no known competing financial interests or personal relationships that could have appeared to influence the work reported in this paper.

**CRediT statement**

**Bing Zhang:** methodology, validation, investigation, writing—original draft, writing—review and editing **Ting Zhang:** methodology, software, writing—review and editing **Jie Pan:** investigation **Tsz Pong Chow:** investigation **Ammar M. Aboalsaud:** resources, investigation **Zhiping Lai:** resources, supervision, project administration, funding acquisition, writing—review and editing **Ping Sheng:** conceptualization, methodology, formal analysis, supervision, writing—original draft, writing—reviewing and editing, project administration, funding acquisition

**Acknowledgements**

P. S. wishes to acknowledge support by the Research Grants Council of Hong Kong, Grant 16308216, and by collaborative grant KAUST18SC01. Z. Lai wishes to acknowledge the KAUST competitive research grant URF/1/3435-01.

## Supplemental Materials

### A. Pore occupation ratio of (3,0) nanotubes

The ZSM-5 unit cell contains 96 silicon atoms and 192 oxygen atoms, with a total mass of 5760 A. U. The TGA result indicates that the carbon weight ratio is 14.3%. A simple calculation shows that the mass of one unit cell of (3,0) CNT is 961 A. U., corresponding to about 80 carbon atoms. Since (3,0) CNT contains 12 carbon atoms inside a single unit cell, we have about ~6.5 unit cells of (3,0) CNT segments inside a single ZSM-5 unit cell. Inside the ZSM-5 unit-cell there are 4 channels that can accommodate (3,0) CNTs, with a total length of ~ 80 Angstrom. Since the length of a (3,0) CNT unit cell is ~4.2 Angstroms, there can be a maximum ~19 unit cells of (3,0) CNTs inside a ZSM-5 unit cell. So the occupation ratio is 6.5/19 ~ 35%

### B. Ab-initio calculations on the soft mode of (3,0)

The observed RBM signal in the Raman spectrum indicate the existence of (3,0) carbon nanotube segments (CNT segments) in the channels of ZSM-5 zeolite structures. Such CNT segment is the smallest CNT ever possible, with a diameter ~3 Angstroms. It is also the most possible candidate structure for filling the ZSM-5 pores, due to the locally one-dimensional geometry and the inner diameter of the ZSM-5 pore. Below we detail the ab-initio calculations involving (3,0) CNT's electron-phonon coupling and the resulting phonon softening/Peierls transition.

We use the density functional theory (DFT) to calculate the structural and the electronic properties of (3,0) CNT, and use the density functional perturbation theory (DFPT) [1] to calculate its lattice dynamics and electron-phonon coupling properties. The latter takes advantage of being able to calculate phonon frequencies at arbitrary q-vectors in the whole Brillouin zone (BZ), which is important in identifying the electron-phonon related instabilities like the Peierls distortion that can cause the metal-insulator transition, as well as the superconductivity transition. A "mixed-basis" pseudo-potential code [2, 3] was used to perform the calculations. We first optimize the geometric structure of (3,0) CNT, then relax the atomic structure to attain the energy minimum, and then calculate the band structure. Results are shown in Fig. S1 and Fig. S2.

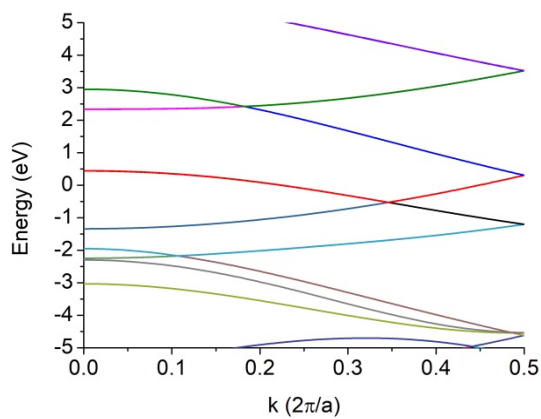

**Figure S1. Calculated band structure of (3,0) CNT. In the spectrum the energy=0 denotes the Fermi level. It is seen that there are a total of 3 bands crossing the Fermi level, which may induce electron-phonon coupling.**

We find in (3,0) CNT that there are three bands that cross the Fermi energy. This result indicates (3,0) CNT to be metallic, and ought to have strong electron-phonon interaction. Therefore in (3,0) CNT system a variety of electron-phonon related physics may occur, like the Peierls distortion and superconductivity. Also the determination of band structure is critical to understand the transport properties as shown in Fig. 4. At high temperatures the sample shows metallic behavior, namely resistance decreases linearly with decreasing temperature. This is consistent with the result that the (3,0) CNT is metallic. At low temperatures, as in one-dimension the Peierls distortion is inevitable, and the occurrence of Peierls distortion, with the accompanying opening of the quasi-gap, explains the metallic to insulator transition in the resistance's temperature dependence.

We have calculated the phonon spectrum of (3,0) CNT at a smearing parameter of electrons to be 0.1 eV and 0.05 eV (corresponds to electron temperature 548 K and 274 K, respectively). Under the first case there is no soft mode in the obtained phonon spectrum, as shown in Fig. S2. We can see that around q=0.45 there is some softness of phonon frequencies, which indicates a strong electron-phonon coupling. From this result we can analyse the Raman active modes depending on their symmetry. We find that for (3,0) CNT the phonon frequency of RBM should be at ~ 817 cm$^{-1}$, which is very close to the observed peak at 805 cm$^{-1}$ as shown in Fig. 3. Therefore the theoretical analysis is consistent with our earlier assumption that in ZSM-5 zeolite the carbon nanostructures are (3,0) CNT segments, forming a 3D network.

We also considered the effect of varying the smearing parameter. The smearing parameter can be considered as an effective electron temperature, and reducing this energy is associated with a cooling down process. While at smearing 0.05 eV, the soft phonon frequency gets more softened and an imaginary phonon frequency is obtained. This means the structure of (3,0) CNT under this temperature is no longer stable, and a Peierls transition occurs above the room temperature $T_p$ ~ 273 K.

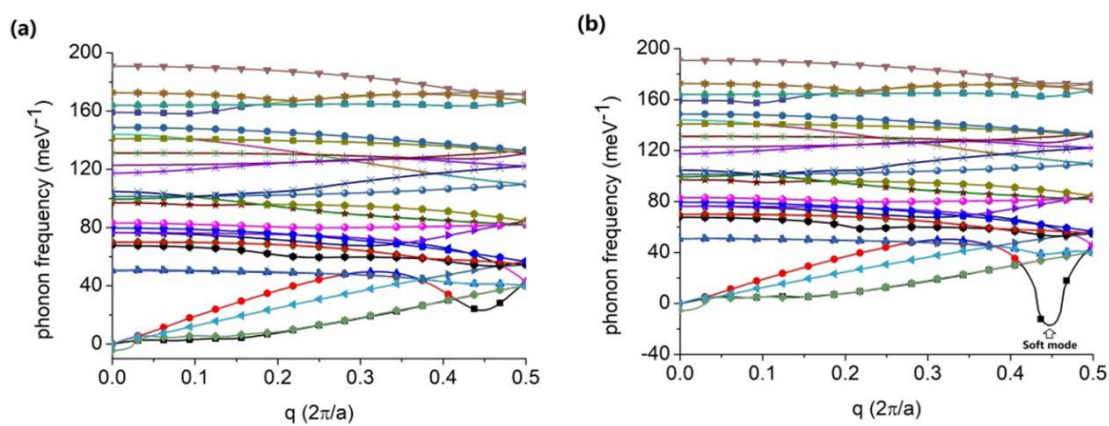

**Figure S2 (a) The phonon spectrum under a smearing broadening of 0.1 eV. (b) The phonon spectrum under a smearing broadening of 0.05 eV. We find that at the 0.05 eV broadening, which corresponds to 273 K, an imaginary phonon frequency occurs, indicating a Peierls transition at this temperature.**

Since an alternative outcome of the electron-phonon interaction is superconductivity, we have also calculated the possible superconducting transition temperature with the Allen-Dynes formula $T_c = 0.833\omega_{\ln}\exp\{-1.04(1+1/\lambda)\}$, with $\omega_{\ln} = \exp\left\{\frac{1}{\lambda}\frac{1}{N_q}\sum_{q,\nu}\lambda_{q,\nu}\ln(\omega_{q,\nu})\right\}$ to be the effective phonon frequency, and $\lambda = \frac{1}{N_q}\sum_{q,\nu}\lambda_{q,\nu}$ is the electron-phonon coupling constant averaged on different phonons labeled by different momentum $q$ and branches. Here $N_q$ is the number of q-points in the calculation, and $\lambda_{q,\nu}$ is the electron-phonon coupling constant for each $q$. Only a smearing parameter value of 0.1 eV result can be used to calculate the superconductivity transition temperature since the superconductivity is competing with the Peierls transition. Once the latter occurs, the system becomes an insulator and superconductivity is excluded. In our case, for (3,0) CNT under 0.1 eV broadening, the averaged electron-phonon coupling constant is ~0.74, the averaged effective phonon frequency $\omega_{\ln}$ is ~38.5 meV, then the estimated superconductivity transition temperature is $T_c$ ~ 2.87 meV or ~ 33 K.

We should keep in mind that both the estimated Peierls transition temperature $T_p$ and the estimated superconductivity transition temperature $T_c$ are calculated in an individual (3,0) CNT. However in the actual case the existence of zeolite channel walls as well as the 3D network structure, may significantly influence the dynamical environment of a (3,0) CNT, and affect related physical properties. The existence of channel walls may set up constrains on the CNT, which can greatly enhance the hardness of the system. Also such constrain is similar to adding pressure on the system, which is favorable to inducing superconductivity if suitable doping is introduced. What's more, the Peierls transition temperature $T_p$ and superconductivity transition temperature $T_c$ are calculated under the random-phase approximation (RPA), while in one-dimensional system it is not a good approximation, and a renormalization group analysis is needed to identify the true ground state of the system. However based on the theoretical calculations up to now we can deduce that in ZSM-5 zeolite channels the carbon nanostructure is very likely to be (3,0) CNTs, evidenced by the agreement between the calculated and observed RBM frequencies and the qualitative agreement between the transport experiments and the calculation results. Our conclusion is therefore in this (3,0)@ZSM-5 system superconductivity is possible to be realized if suitable doping is accomplished, since an optimized free-electron concentration can significantly enhance the screening of Coulomb repulsion in one-dimension, which is the main reason of a Peierls distortion state.

**References for Supplemental Materials**